\documentclass[12pt]{iopart}
\pdfoutput=1
\expandafter\let\csname equation*\endcsname\relax
\expandafter\let\csname endequation*\endcsname\relax

\usepackage{graphicx}
\usepackage{amssymb}
\usepackage{amsmath}
\usepackage{bm}
\usepackage{url}
\usepackage{epstopdf}
\usepackage{iopams}  

\begin{document}
\title{Free Electron Lasers using `Beam by Design'}
\author{J. R. Henderson$^{1,2}$, L.T. Campbell$^{1,2}$ and B.W.J. M$^{\mathrm c}$Neil$^{1}$}
\address{1 SUPA, Department of Physics, University of Strathclyde, Glasgow, G4 0NG, UK}
\address{2 ASTeC, STFC Daresbury Laboratory and Cockcroft Institute, Warrington, WA4 4AD, UK}
\eads{\mailto{j.r.henderson@strath.ac.uk}, \mailto{lawrence.campbell@strath.ac.uk}, \mailto{b.w.j.mcneil@strath.ac.uk}}
\submitto{\NJP}

\maketitle

\begin{abstract}
Several methods have been proposed in the literature to improve Free Electron Laser output by transforming the electron phase-space before entering the FEL interaction region.  By utilising `beam by design'  with novel undulators and other beam changing elements, the operating capability of FELs may be further usefully extended. This paper introduces  two new such methods to improve output from electron pulses with large energy spreads and the results of simulations of these methods in the 1D limit are presented. Both methods predict orders of magnitude improvements to output radiation powers.
\end{abstract}

\section{Introduction}
The Free Electron Laser (FEL) is an important scientific research tool that uses a relativistic electron beam to generate coherent radiation from the microwave through to the hard X-ray. At shorter wavelengths into the X-ray, this is unlocking many new areas of science in diverse fields such as: Warm-Dense matter studies~\cite{Vinko}; short pulse protein diffraction~\cite{Barends} and medicine/surgery~\cite{Edwards}. 
Current X-ray FELs~\cite{lcls,sacla} and those under construction~\cite{McNeil xray}, are unique laboratory sources of high power coherent X-rays. They are driven by electron beams generated from Radio-Frequency linear accelerators, which can be up to a few kilometres long.  

Many ideas are now being proposed to enhance and improve FEL output, towards shorter wavelengths, shorter output pulse durations, improved temporal coherence~\cite{beam by design} and multi-colour operation~\cite{2colour}.  These improvements extend the original  high-gain FEL design where the  electron beam from an accelerator is simply injected into a long undulator where the collective FEL interaction generates coherent output. The new methods  rely upon manipulation of the electron beam in phase-space, using laser modulators and magnetic chicanes, either prior to injection into the FEL, or sequentially  along the undulator as the FEL interaction progresses.

Proposals also exist to reduce the overall lengths of FEL facilities by replacing the RF-linacs with plasma-wakefield accelerators~\cite{Couprie,Nakajima}. These accelerators have large accelerating gradients about $10^3-10^4$ times larger than RF-linacs. However, the electron bunches generated so far are limited by a relatively large energy spread which  inhibits any useful FEL interaction.
As with the above proposed enhancements, methods that manipulate the electron beams have been proposed that may help mitigate the detrimental effects of energy spread. 
These include stretching the beam longitudinally before injection into the FEL to reduce the localised energy spread~\cite{Maier}, or transversely dispersing the electron beam to give a correlated transverse energy distribution and then matching this into a transverse gradient undulator~\cite{Schroeder}.

Using a combination of modulators and chicanes, it is also possible to fourier-compose electron pulses of simple geometric shapes in longitudinal electron beam phase space e.g. rectangular, triangular, and sawtooth~\cite{rf-function beams}. Such waveform synthesis of the electron beam can also be utilised to generate phase-correlated harmonic beam structures that can then perform analogous waveform synthesis of the coherent light emission from the beam structures. 

The electron beam parameters and manipulations described above can be very difficult, if not impossible, to model using conventional FEL simulation codes, which average the FEL interaction over a resonant radiation wavelength limiting both the radiation bandwidth that can be modelled and the range of electron energies, correlated or uncorrelated, within the beam. 

In this paper the un-averaged FEL simulation code PUFFIN~\cite{Campbell PUFFIN} is used to simulate potentially useful electron beam undulator emission that would not be possible using conventional averaged FEL simulation codes. 

Firstly, a new method using electron beam phase-space manipulation is investigated, that may allow a FEL to operate with larger electron beam energy spreads which, for example, may assist the drive towards plasma-accelerator driven FELs. The method constructs a series of energy-chirped electron pulses (beamlets), each of different mean energy,  vertically stacked  in energy in phase space. The localised, or `slice', energy spread of each beamlet is smaller than the original, unmodified beam from which the beamlets are constructed. Previous work has  used multiple beams generated individually by a photocathode illuminated by multiple light pulses to generate different colour pulses from a FEL~\cite{italy1}. Here, however, the beamlets are generated from a single electron pulse.

Secondly, an example is presented of what may be possible using fourier-synthesised electron beams~\cite{rf-function beams}. This is the first simulation of the output from such waveforms in a FEL-type system.  A fourier-synthesised electron pulse with a rectangular wave structure in phase space is used to generate radiation in a series of undulator-chicane modules similar to those used in a mode-locked FEL amplifier~\cite{mode-locked 1}. The `discontinuous' regions of the square electron pulse form larger current regions that can emit significant coherent spontaneous emission (see e.g.~\cite{McNeil SACSE}).  This coherent emission is periodically superimposed using a sequence of undulator-chicane modules and is shown to be able to generate significant output powers. This cannot strictly be called a FEL as little FEL interaction takes place.  

The methods simulated here are clearly not to be considered as specific FEL design proposals, rather they are intended to demonstrate future possibilities and potential as electron beam generation advances beyond that of a simple linear beam model.

\section{Beamlets}
\subsection{Beamlets - Description of Method}
In the Free Electron Laser (FEL), a relativistic electron beam of mean electron energy $\gamma_r m_ec^2$ amplifies radiation in an undulator of period  $\lambda_u$ and rms magnetic field strength $B_u$. The resonant radiation wavelength amplified is given by $\lambda_r=\lambda_u(1+a_u^2)/2\gamma_r^2$. The high-gain amplification process is characterised by the gain length $l_g$, where an initial radiation power $P_0$ is amplified exponentially as a function of the distance $z$ through the undulator as $P(z) = P_0 \exp(\sqrt{3} z/{l_g})$~\cite{bnp}. With an electron beam energy of $\gamma_r$, the gain length may be written, neglecting radiation diffraction and for no electron beam energy spread $\sigma_\gamma=0$, as:
\begin{align}
l_{g} = \frac{\lambda_u}{4\pi\rho}=\frac{1}{2k_u\rho},
\end{align}
where: $k_u=2\pi/\lambda_u$, 
\begin{align}
\rho = \frac{1}{\gamma_r}\left( \frac{\bar{a}_u \omega_p}{4ck_u}  \right)^{2/3}  \propto I_{pk}^{1/3},
\end{align}
is the is the FEL (or Pierce) parameter, $\bar{a}_u \propto B_u k_u$ is the undulator parameter,  $\omega_p$ is the peak (non-relativistic) plasma frequency of the beam, and $I_{pk}$ is the peak current. 
For good amplification, the electron beam energy spread $\sigma_\gamma$ must satisfy the `cold beam' limit of:
\begin{align}
\sigma_p = \frac{\sigma_\gamma}{\rho \gamma_r} \ll 1. \label{espcond}
\end{align}
Optimal FEL gain is seen to occur when $I_{pk}$ is maximised and  $\sigma_\gamma$ minimised. The method described below uses electron beam phase space manipulation to modify both of these parameters in an attempt to improve the FEL output potential of beams with large energy spreads ($\sigma_p \gtrsim 1$). 

The method first generates a series of energy chirped beamlets stacked vertically in longitudinal phase space before they are injected into the FEL amplifier.  As the FEL interaction occurs within the undulator further manipulation is required to ensure the radiation interaction with the chirped electron beamlets maintains a resonant interaction. 

\begin{figure}
\centerline{\includegraphics[width=140mm,height=110mm]{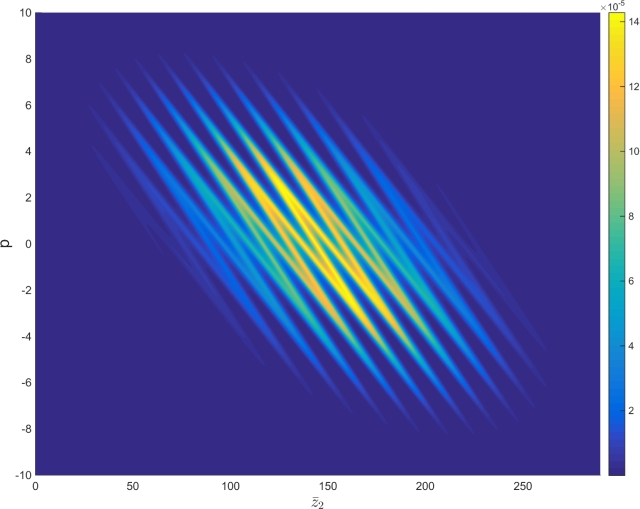}}
\caption{The scaled longitudinal electron beam phase space distribution function  given by equation~(\ref{beamlet dist function}) $f(\bar{z}_2,p)$ (using $p$ rather than $\gamma$) after transformation by a beam modulator and dispersive chicane. } \label{figure1}
\end{figure}

In the first stage before injection into the FEL, the electron beam is passed through a modulating undulator and dispersive chicane, resulting in the beam phase space shown in figure \ref{figure1}. This phase space is similar to the first modulator-chicane section used in the Echo Enhanced Harmonic Gain method~\cite{Stupakov}. The modulator-chicane sections perform the following consecutive transforms on the electron beam phase space coordinates:
\begin{align}
\gamma = \gamma_{0} - \Delta \gamma   \sin \left( \frac{ \bar{z}_{20}}{2 \rho n} + \phi \right) \label{beammod} \\
\bar{z}_{2} = \bar{z}_{20} - 2 D \left( \frac{\gamma - \gamma_r }{\gamma_r} \right) \label{beamdisp},
\end{align}
where the subscript $0$ denotes the initial, untransformed coordinates, $\bar{z}_2 = (ct - z)/l_c$ is the coordinate in a window travelling at the speed of light scaled with respect to the cooperation length $l_c=\lambda_r/4\pi\rho$ of the FEL interaction, $\Delta\gamma$ is the energy modulation amplitude, $n =  \lambda_1/\lambda_r$  is the modulation period scaled with respect to the resonant wavelength and $D =k_r \rho R_{56}$ is the scaled dispersive strength of the chicane. With this scaling, a resonant electron of energy $\gamma_r$ will fall behind a resonant radiation wavefront a distance $l_c$ on propagating one gain length $l_g$ through the undulator~\cite{bmp}.

It has been observed that in regimes where large dispersion is applied that the noise statistics of the macroparticles that simulate the electrons in the dispersed beam can become incorrect. This occurs as the beam sampling in $\bar{z}_2$ is transformed into the $\gamma$ dimension when rotated in phase space, and vice-versa. To ensure the  correct noise is modelled, the functional form of the final electron beam phase space is used to initialize the beam before application of the noise algorithm~\cite{noise} and simulation using Puffin.

A gaussian distribution for both dimensions of the initial beam phase space is assumed:
\begin{equation}
f(\bar{z}_2,\gamma) =  \frac{1}{2 \pi \sigma_\gamma \sigma_{\bar{z}_2}} \exp\left[-\frac{ (\gamma - \gamma_r)^2}{2 \sigma_\gamma^2} \right] \exp\left[-\frac{(\bar{z}_2 - \bar{z}_c)^2}{2 \sigma_{\bar{z}_2}^2} \right],  \label{initial dist function}
\end{equation}
where: $\bar{z}_c$ is the electron pulse centre and $\sigma_{\gamma,\bar{z}_2}$ are the standard deviations in $\gamma$ and $\bar{z}_2$ respectively. 

By applying similar modulation and dispersive  transforms to those outlined in~\cite{Stupakov}, the final beam distribution function obtained is:
\begin{align}
f(\bar{z}_2,\gamma) =  \frac{1}{2 \pi \sigma_\gamma \sigma_{\bar{z}_2}} & \exp\left[-\frac{1}{2 \sigma_\gamma^2}  \left( \gamma + \Delta \gamma   \sin \bigg[ \frac{ 1}{2 \rho n}  \bigg(   \bar{z}_{2}   \right. \right. 
 \left. \left.  + 2 D \left( \frac{\gamma - \gamma_r }{\gamma_r} \right)  \bigg) + \phi  \bigg] - \gamma_r\right)^2 \right] \nonumber \\ 
& \times \exp\left[-\frac{1}{2 \sigma_{\bar{z}_2}^2}  \left( \bar{z}_{2} + 2 D \left( \frac{\gamma - \gamma_r }{\gamma_r} \right) - \bar{z}_c \right)^2 \right] \label{beamlet dist function}.
\end{align}

Figure \ref{figure1} plots the scaled longitudinal phase space distribution function of the electrons after the modulation-dispersive section and before injection into the FEL undulator using the scaled energy parameter $p_j  = (\gamma_j - \gamma_r)/\rho \gamma_r$ with the following parameters:  $\Delta \gamma = 0.04 \gamma_r$, $D=268.51$, $n = 68$, $\phi = 0$, $\sigma_\gamma = 2 \rho \gamma_r$ (or $\sigma_p=2$), $\gamma_r = 1200$, $\rho = 1.6\times10^{-2}$ and $\sigma_{\bar{z}_2} = 28.97$.  The modulation and dispersion of the beam is seen to create a stacked structure of energy chirped `beamlets', slice sections  of which are seen to have an energy spread which is reduced from the initial un-transformed beam with $\sigma_p=2$. Under certain conditions, each beamlet may then emit and amplify radiation independently of the other beamlets. The combined output from each of the beamlets may then give improved radiation output over the un-transformed beam.

\begin{figure}
\centerline{\includegraphics[width=140mm,height=110mm]{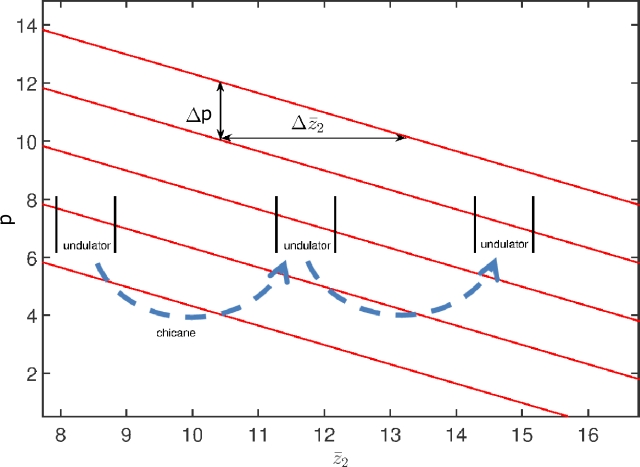}}
\caption{Scaled longitudinal phase space of the electrons for the simplified beamlet model. For a given energy the beamlets are separated spatially by $\Delta \bar{z}_2$ and for a given $\bar{z}_2$ the beamlets are separated by $\Delta p=\Delta \gamma/\rho\gamma_r$. A chicane delay of the electrons corresponds to a positive shift in $\bar{z}_2$. A series of chicanes slip the electrons forward in $\bar{z}_2$ so that they interact with the same resonant wavelength as emitted by the previous beamlet.} \label{figure2}
\end{figure}

To illustrate how the method functions in the FEL undulator, a simplified version of the beamlet phase space is shown in figure~\ref{figure2}, which consists of a series of chirped, zero energy spread, electron beamlets of different mean energies stacked in phase space. 
The chirp causes the radiation from one section of the chirped beam to drift out of resonance as it propagates into electrons which are resonant at a different wavelength. This impedes the FEL gain process. This effect may be successfully counteracted by using an appropriate undulator tapering to maintain the electron-radiation resonance~\cite{Saldin chirp paper}. (These results have been reproduced using the simulation methods used here and are in very good agreement~\cite{fel14}.) Here, a different approach is demonstrated using a periodic series of undulator-chicane modules with multiple beamlets.  The  beamlets are periodically delayed by the chicanes so as to maintain a resonant interaction with the radiation generated by electrons of the same energy from the other beamlets.  (Simulations using this method on the simple beamlet structure of figure~\ref{figure2} have been performed and presented elsewhere~\cite{fel14}.) In the electron beam frame therefore, the radiation is passed from beamlet to beamlet so that it always interacts with electrons of a similar energy so maintaining a resonant interaction and giving an improved FEL interaction. This is achieved by making the slippage of a radiation wavefront through the electrons in each undulator-chicane module equal to the spatial separation of the beamlets. The enhanced slippage can also be expected to result in the generation of a series of modes in the radiation spectrum similar to that of~\cite{mode-locked 1} which demonstrated that an undulator-chicane lattice will amplify side-band radiation modes that are separated by:
\begin{equation}
\Delta \omega / \omega_r = 4 \pi \rho /  \bar{s}, \label{mode-locking-freq}
\end{equation}
where $\bar{s}$ is the slippage length in scaled units of $\bar{z}_2$ in one undulator-chicane module~\cite{mode-locked 1}.

\begin{figure}
\centerline{\includegraphics[width=140mm,height=110mm]{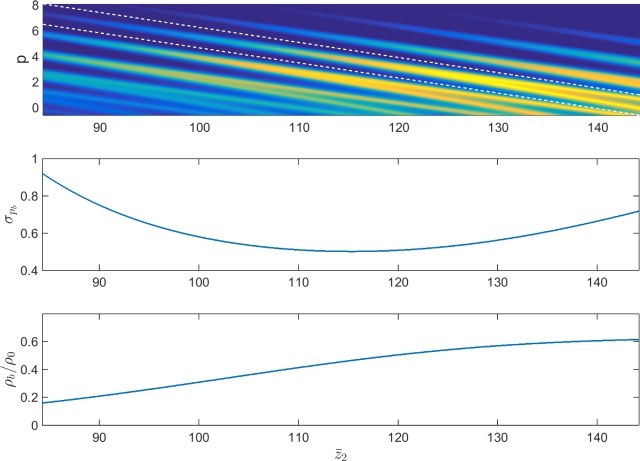}}
\caption{Top panel: Detail of the higher energy beamlet phase space distribution function of equation~\ref{beamlet dist function} with a single beamlet delineated by white dashed lines. The values of the scaled energy spread $\sigma_{p_b}$ (middle) and $\rho_b/\rho_0$ (bottom) were calculated for the single beamlet as a function of $\bar{z}_2$. Towards the pulse head ($\bar{z}_2<105$) the electron pulse is diffuse with a larger energy spread $\sigma_{p_b}$ and smaller $\rho_b$.  Nearer the centre of the pulse ($105<\bar{z}_2<125$), the scaled energy spread decreases as the local density, and $\rho_b$ increase.  However, further towards the pulse centre $\bar{z}_2>125$ the energy spread increases further as the beamlet spilts into two identifiably separate beamlets, while the value of $\rho_b$ tends towards a more constant value.  The condition for lasing of the beamlet of $\sigma_{p_b} < 1$ is seen to be satisfied within this the head of the pulse (and is also satisfied at the lower energy beamlets of the tail).  The  energy separation between beamlets is also seen to satisfy condition~(\ref{sepcond})  so that each beamlet can lase independently.  The energy separation between beamlets does not change significantly with $\bar{z}_2$, as neither does the longitudinal separation of beamlet regions with the same energy. Towards the centre of the pulse however, the beamlet structures have a more complicated phase space structure.} \label{beamlet head}
\end{figure}

\begin{figure}
\centerline{\includegraphics[width=140mm,height=110mm]{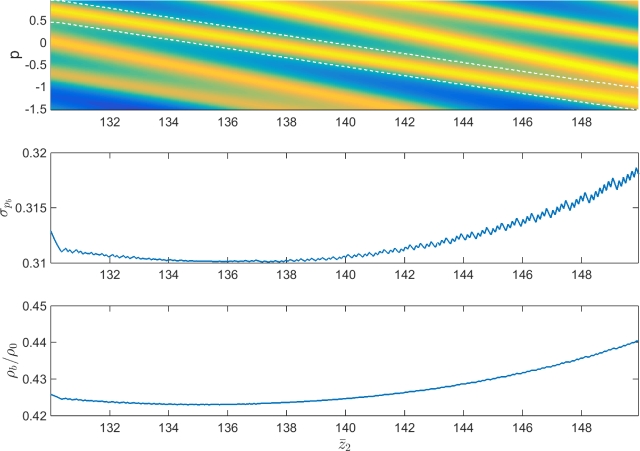}}
\caption{As figure~\ref{beamlet head}, but around the centre of the electron pulse about the mean pulse energy. The beamlets are seen to `spilt' into two separate beamlets. While the scaled energy spread requirement for lasing $\sigma_{p_b} < 1$, is satisfied, the beamlet energy separation condition~(\ref{sepcond}) is only satisfied for a small region of beamlets about the pulse centre.  The beamlets are therefore unlikely lase  independently with non-overlapping bandwidths, so that the effective energy spread for the interaction is increased, decreasing the ability of achieving significant FEL lasing.} \label{beamlet centre}
\end{figure}

The FEL parameter $\rho \propto I_{pk}^{1/3}$, where $I_{pk}$ is the electron pulse peak current, and is a measure of FEL efficiency.  When considering individual beamlets a FEL parameter may also be defined for each beamlet: $\rho_b \propto I_{b}^{1/3}$ where  $I_b$ is localised (slice) current of  the beamlet.  (Note that as the beamlet energy is chirped, the mean pulse energy $\gamma_r$, is used in the definition of $\rho_b$.) Other beamlet parameters are also defined as  $p_{b} = (\gamma_j - \gamma_r)/\rho_b\gamma_r $ and a beamlet scaled slice energy $\sigma_{p_{b}}$.  For a beamlet to lase independently its slice energy spread must then satisfy:
\begin{align}
\sigma_{p_b} = \frac{\sigma_{\gamma_b}}{\rho_b\gamma_r} <  1. \label{espcondb}
\end{align}
(Note here, that the mean pulse energy $\gamma_r$ is used in the definition of $\sigma_{p_b}$ rather than a local `slice' value $\gamma_b$. This can be considered a reasonable approximation for the inequality~(\ref{espcondb}), so long as $\gamma_b$ does not differ significantly from $\gamma_r$.)
The beamlet slice energy spread $\sigma_{p_{b}}$ and instantaneous fractional FEL parameter $\rho_b/\rho_0$, where $\rho_0$ is the FEL parameter of the un-transformed beam,  can be calculated and are shown in figures~\ref{beamlet head} and \ref{beamlet centre} towards the higher energy and mid-sections of the electron pulse respectively. 

The energy spread condition for FEL lasing of equation~(\ref{espcondb}) may be used with the FEL radiation bandwidth saturation $\Delta \omega / \omega_r \approx  2 \rho$~\cite{bprl} to define the minimum energy separation $\Delta \gamma$ of the beamlets so that the gain bandwidths of each beamlet do not overlap: 
\begin{align}
\frac{\Delta \gamma}{\rho_b\gamma_r} \gtrsim 2 \label{sepcond}.
\end{align}
At the centre of the electron pulse the beamlets  split  into pairs~\cite{JamesHenderson}, i.e. two per half modulation period, while for the electron pulse higher and lower energies, formed by the modulation extrema, the beamlet pairs merge into single beamlets as seen in figures~\ref{beamlet head} and \ref{beamlet centre}.

Both the energy spread condition~(\ref{espcondb}) and beamlet separation condition~(\ref{sepcond}) are seen in figure~\ref{beamlet head} to be satisfied for the  higher energy regions of the beamlets. (These conditions are also satisfied at the lower energy regions of the beamlets, but are not shown.) However, the condition placed on the beamlets' energy separation~(\ref{sepcond}) is not always satisfied at the pulse centre where the beamlets are formed in pairs, as seen in  figure~\ref{beamlet centre}.    Hence, the undulator-chicane slippage length is set equal to the beamlet separation for the higher and lower energy regions of the pulse where the energy separation of the beamlets is approximately a constant.

\begin{figure}
\centerline{\includegraphics[width=140mm,height=110mm]{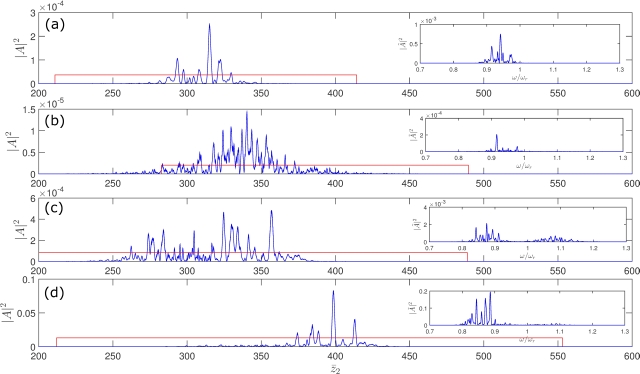}}
\caption{A comparison of the scaled radiation temporal power and spectral power (insets) for an un-transformed electron pulse (panels a and b) and transformed pulse of beamlets (panels c and d), when propagated through an simple undulator and an undulator-chicane lattice respectively and interaction length of $\bar{z}\approx 30$.  The (red) box shows the position of the electron pulse relative to the radiation (the head of the pulse is to the left.) Note the different lengths of the electron pulses due to differing dispersive effects of the chicanes. The beamlets propagating through a simple undulator (panel c) is seen to give a small improvement to the output  from the un-transformed beam through both an simple undulator and an undulator-chicane system (a and b respectively.) The improvement in output from the beamlets is increased significantly when they are propagated through the matched  undulator-chicane lattice as shown in panel d. The undulator-chicane lattice amplifies side-band radiation modes generated by the undulator-chicane modules and are separated by $\Delta \omega = 0.0147$  as seen from the panel d inset and in agreement with the mode-spacing relation of~(\ref{mode-locking-freq}).  For all results shown in this figure the radiation field has been filtered about the resonant frequency  $0.5 < \omega/\omega_r < 1.5$ to eliminate low frequency coherent spontaneous emission.} \label{beamlets results}
\end{figure}

Results of a FEL interaction using an un-transformed (no beamlets) pulse with large energy spread $\sigma_p=2$ and of the transformed (beamlet) pulse are shown in figure~\ref{beamlets results}. As expected, for the pulse without beamlets and the large energy spread, only small scaled peak powers of $|A|^2 \sim 10^{-4}$ are observed in the simulation. However, for the transformed pulse with beamlets that have smaller energy spread, $\sigma_{p_b}<1$, and that are matched to the undulator-chicane modules, powers 2-3 orders of magnitude greater are observed. For the modulation period of $68 \lambda_r$ used here ($n=68$), matching was achieved using undulator modules of $20$ periods and isochronous chicane slippages of $48 \lambda_r$.
It is seen that the FEL lasing is greater for the lower energy beamlets of the pulse around $\bar{z}_2\sim 400$.  This preferential FEL interaction and amplification of the lower frequency is consistent with the scaling of the FEL parameter $\rho \propto \gamma^{-1}$ which gives greater values and so strength of interaction, for lower beam energies. In the simulations here, the gain length of the higher to lower energy beamlets is up to $\sim$50\% larger. 
Evidence of the modal structure in the spectrum is also observed in the scaled power spectrum (inset), consistent with the undulator-chicane system which from~(\ref{mode-locking-freq}) gives a mode spacing of $\Delta \omega = 0.0147$.

\begin{figure}
\centerline{\includegraphics[width=140mm,height=110mm]{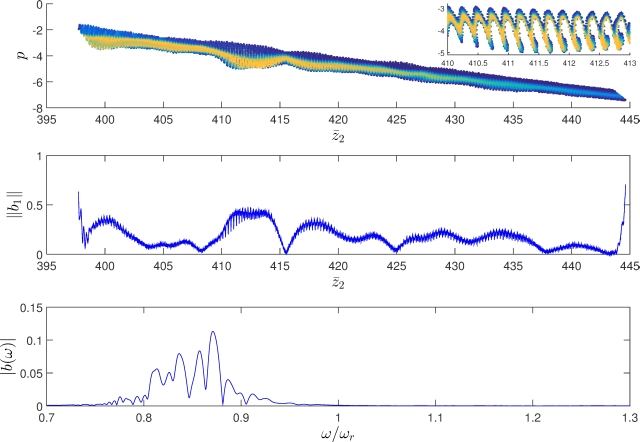}}
\caption{Electron bunching in a lower energy beamlet at $\bar{z}\approx 30$. The top panel plots the charge-weighted electron phase space distribution ; the middle plots the bunching parameter of the beamlet at the fundamental radiation frequency, and the bottom plots the bunching spectrum of the beamlet.} \label{beamlet bunching low}
\end{figure}

\begin{figure}
\centerline{\includegraphics[width=140mm,height=110mm]{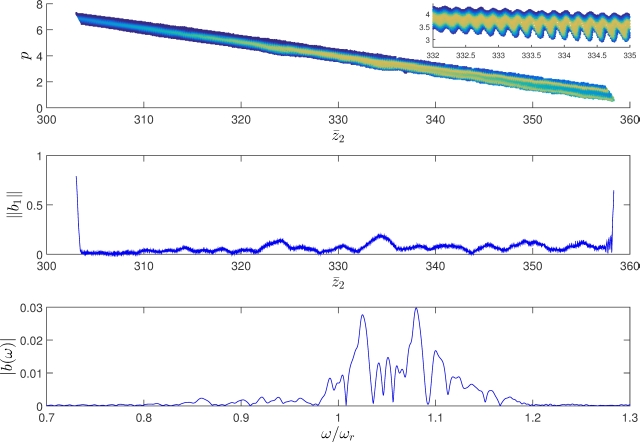}}
\caption{Electron bunching in a higher energy beamlet at $\bar{z}\approx 30$. The top panel plots the charge-weighted electron phase space distribution; the middle plots the bunching parameter of the beamlet at the fundamental radiation frequency; and the bottom plots the bunching spectrum of the beamlet.} \label{beamlet bunching high}
\end{figure}

Significant bunching of the electrons in one of the lower energy beamlets, with a mean value of scaled energy $ <p>  = (< \gamma > - \gamma_r)/ \rho \gamma_r \approx -5$, is also observed as shown in figure~\ref{beamlet bunching low}. Note from the lower plot for the spectrum that the electrons are bunched at a lower frequency $\omega/\omega_r\approx 0.85$ than the mean resonant frequency of the electron pulse. This frequency shift from resonance is consistent with the lower mean energy of the electrons as $\Delta\omega/\omega_r\approx 2\rho <p>=0.16$ and is in agreement with the radiation frequency spectrum of figure~\ref{beamlets results}.
Electron bunching is also observed in a higher energy beamlet of mean scaled energy $<p>\approx 4$, shown in figure~\ref{beamlet bunching high}. Here, the bunching is seen to be at a less advanced stage, but can be expected to reach saturation on further propagation through the undulater-chicane lattice.

\section{Fourier Synthesised Electron Beams}
Further types of phase-space transformation of an electron pulse prior to generating radiation have been proposed and called `beam-by-design'~\cite{beam by design}. An example is investigated here to demonstrate the potential of such beam transformation prior to injection into the FEL and the subsequent transformation in the FEL emission stage using a series of undulator-chicane modules.
An electron pulse consisting of a series rectangular shaped distributions in phase-space can be generated~\cite{rf-function beams} and contains a periodic series of current `spikes'.  These current spikes are a source of coherent spontaneous emission which may, through a series of periodic superpositions enabled by chicanes, generate significant radiation output from an undulator-chicane lattice. We note that other methods can generate similar beam structures, e.g. the E-SASE approach~\cite{esase}, however the methods of~\cite{rf-function beams} are used here to demonstrate the types of more exotic interaction that may be modelled using non-averaged simulation codes such as PUFFIN.

\subsection{The Model - Coherent Emission from Rectangular Electron Pulses}
A new approach to produce so-called `RF-function' electron beams was introduced in~\cite{rf-function beams}.  
An RF-function generator produces a series of repeated wave forms by combining sine-waves of different frequencies as in a Fourier series.  
In a similar fashion, an electron pulse can be created with a phase-space that consists of repeated `waveforms' by modulation the electron beam using a series of seeded undulator modulators using different seed wavelengths, amplitudes and phases. 
Following the notation of~\cite{rf-function beams}, here a rectangular beam shape in phase space using a triple modulator-chicane lattice is synthesized and subsequent radiation generation following injection into an undulator chicane-lattice is modelled using PUFFIN.

While in~\cite{rf-function beams} an infinity long electron beam was assumed, here, a finite electron pulse with an initial Gaussian distribution in both $\bar{z}_2$ and  $\gamma$ is assumed, as given by equation~(\ref{initial dist function}).  As detailed in the Appendix, the same Fourier synthesis as outlined in~\cite{rf-function beams} is applied using the beam modulation transforms given by equation~(\ref{beammod}) and the energy dispersion transforms of equation~(\ref{beamdisp}).
 
In electron phase-space, the vertical segments of the rectangular waveform generate regions of enhanced current, albeit with a larger energy spread.  
Each period therefore contains two current `spikes' which can generate significant coherent spontaneous emission when their width is of a similar scale to a resonant wavelength~\cite{McNeil SACSE}. 
However, due to electron beam dispersion in the undulator, the sharpness of the current spikes reduce on propagation, resulting in diminishing coherent emission. 
This dispersion of the current spikes may be compensated for by the use of chicane systems with a negative dispersion to allow for more prolonged coherent emission. 
The design of chicane delay systems with negative dispersion have been previously designed and tested as part of an accelerator lattice~\cite{James Jones} and are also necessary for generating the RF-function beam shapes~\cite{beam by design,rf-function beams}.  
If the slippage per undulator-chicane module is also made equal to the current spike separation, then the radiation is propagated from spike to spike and, if correctly phased, can  facilitate the constructive interference of the coherent emission from each current spike in each new undulator module.

\subsection{Results - Coherent Emission from Rectangular Electron Pulses}
The following simulations use the same electron pulse parameters as the previous section, i.e., the electron pulse's large energy spread is prohibitive to FEL gain.
The phase-space distribution of the electron beam for the rectangular waveform was constructed from the analysis of the Appendix for three undulator-chicane modules using the following  parameters in  $[\Delta \gamma,D]$: $[\Delta \gamma_1 = 10 \sigma_\gamma; D_1 = n_1 \rho \gamma_r \sqrt{3} \pi / (2 \Delta  \gamma_1)]$; $[\Delta \gamma_{2}= \Delta \gamma_1 / 4; D_2 = -3 D_1]$; $[\Delta \gamma_3 = \Delta \gamma_2/16; D_3 = -3 D_2 /4=9 D_1/4]$, with $n_{1,2,3} = 20$,   $\phi_{1,2} = 0$ and $\phi_{3} = \pi$. 

\begin{figure}
\centerline{\includegraphics[width=180mm,height=100mm]{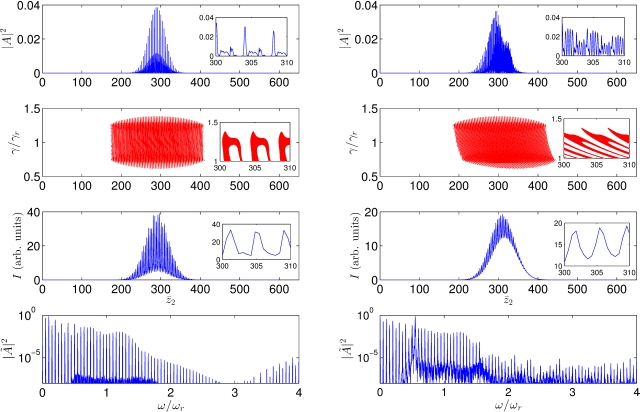}}
\caption{The evolution of a rectangular electron beam in an undulator showing top-to-bottom, the scaled radiation power $|A|^2$ as a function of $\bar{z}_2$, electron phase space ($\gamma/\gamma_r, \bar{z}_2$) with detail inset, the scaled electron current as a function of $\bar{z}_2$ and the logarithm of the scaled radiation power spectrum $|\tilde{A}|^2$ as a function of the scaled frequency $\omega/\omega_r$. The series on the left plot the output for a scaled distance through the undulator of $\bar{z}\approx 1$ and on the right for $\bar{z}\approx 20.1$ 
An electron pulse with an initially large energy spread has been transformed into an electron pulse that contains a number of rectangular waveforms (see second plot on the left).  The electron pulse structure now contains a series of current spikes of spacing  $\Delta \bar{z}_2\approx 4$.  When this electron pulse is passes through an undulator each current spike acts as a source of coherent spontaneous emission.  The radiation spectrum (bottom panels) show a broad bandwith modal structure with modes separation $\Delta \omega/\omega_r \approx 0.05$.  As the electron pulse propagates along the undulator, the rectangular waveforms will disperse, and increase the current spike widths and reduce current spike amplitudes.  As the current spikes' 'sharpness' decrease the coherent radiation produced by the current spikes will decrease.  Because of this no amplification is seen when passing such an electron pulse through a long undulator, as shown in the r.h.s. of this figure.} 
\label{new rect figure 1}
\end{figure}

The initial current profile of the electron pulse  contains a series of current spikes at half the modulation period corresponding to $10$ resonant radiation wavelengths or $10 \times 4 \pi \rho\approx 2$ in units of $\bar{z}_2$. 
On injection into an undulator, these spikes act as a periodic series of phase correlated coherent emitters which, for a relatively short interaction length of $\bar{z} \approx 1)$, generate a broad modal radiation spectrum as seen figure~\ref{new rect figure 1}.  
However, it is seen that alternate current spikes have dispersed to leave a series of more prominent current spikes at twice the initial spacing of $\Delta \bar{z}_2\approx 4$.  
This is reflected in the temporal separation of the larger radiation spikes separated by $\Delta \bar{z}_2 \approx 4$). This also agrees with  to the spectrum in which a series of modes are generated with separation, from equation~(\ref{mode-locking-freq}), of $\Delta \omega /\omega_r\approx 0.05$ about the resonant frequency.

On propagating further through the interaction region to larger values of $\bar{z}\approx 20$, the right hand panels of figure~\ref{new rect figure 1} show that the energy modulation of the rectangular electron beam causes the electron beam to disperse in the undulator degrading the visibility of the current spikes and so decreasing the coherent spontaneous emission generated.  Clearly, these dispersive effects mean that there is no benefit in increasing the interaction length over that of $\bar{z} = 20 \pi \rho \approx 1$.

\begin{figure}
\centerline{\includegraphics[width=140mm,height=100mm]{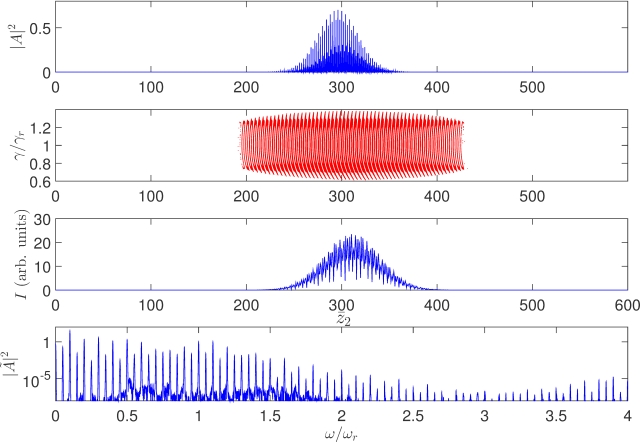}}
\caption{By using  chicanes with a negative dispersion, the undulator dispersion of the rectangular sections of the electron beam can be partially compensated for as seen here for $\bar{z}\approx 10$.  In doing so the electron pulse can continue to emit coherent emission in each  undulator module.  Here, each undulator module has 5 periods and each chicane delays the electron pulse by approximately 5 resonant periods, to match the current spike separation.    Note that there is small FEL interaction as evidence by electron microbunching (not shown).
\label{CHICANES}}
\end{figure}

By using chicanes with a negative dispersion it is possible to partially compensate for the undulator dispersion and maintain a spiked current  profile that can continue to emit CSE over a larger number of modules.  An example of this is shown in figure~\ref{CHICANES} were  the chicane dispersion is set equal the negative of the undulator dispersion, i.e. $D =  - \bar{l}$~\cite{fel14}. The total undulator-chicane slippage for the radiation was again set equal to the current-spike separation, $\bar{s} =  10 \times 4 \pi \rho$.  For this case, undulator-chicane modules of 5 undulator periods and 5 chicane slippage periods were used. In this way, the CSE from successive undulator-chicane modules superimpose and constructively interfere increasing the radiation power emitted. 

\begin{figure}
\centerline{\includegraphics[width=140mm,height=100mm]{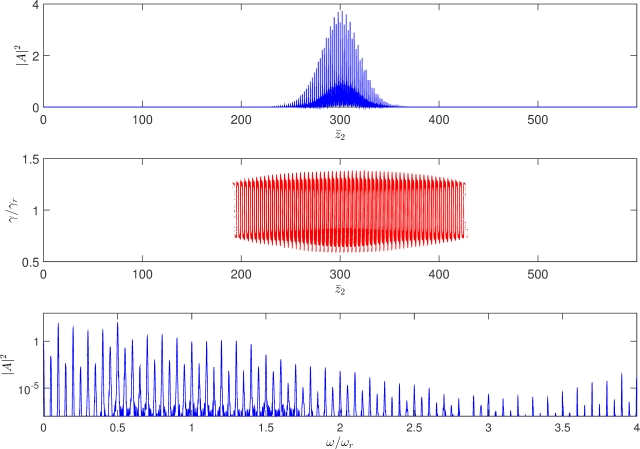}}
\caption{As figure~\ref{CHICANES}, but now using an optimised chicane which maintains the rectangular waveform electron pulse structure in phase space as it propagates through the undulator-chicane lattice.  The rectangular electron waveform emits  coherent radiation in each new undulator module which constructively interferes with the radiation in subsequent undulator modules.
\label{optCHICANES}}
\end{figure}

However,  the radiation fields from each undulator-chicane do not superimpose coherently and the radiation energy is seen (not shown) to scale approximately as the number of  undulator-chicane modules - a phase-matched coherent superposition would give a radiation energy which scales as the square of the number of  undulator-chicane modules. 
The reason for this non-coherent superposition is that the dispersion of the large energy modulated beam in the undulators  cannot be perfectly compensated for by the negative dispersion in the chicanes. (Phase space dispersion of electrons in the undulator is due to differences in the axial speed $v_z$, while electron dispersion in the chicanes is due to differences in the electron energy, $\gamma$.) This is observed from the slight `bowing' of the rectangular  structure of the electrons in phase space in figure~\ref{CHICANES}. Two possible methods to improve this are to reduce the initial energy modulation of the rectangular wave (the results here are for a relatively large energy modulation) or to use a (hypothetical) optimised chicane design which has a non-linear dispersive strength as a function of $\gamma$. Here the latter is used and the results shown in figure~\ref{optCHICANES}. Now, the bowing of the rectangular structure of the electrons in phase space is seen to be removed and the power of the radiation increased. The coherent radiation from each undulator-chicane module is now phase matched and is superimposing coherently after each module. The radiation energy is now also observed to increase in proportion to the square of the number of modules.

\begin{figure}
\centerline{\includegraphics[width=120mm,height=100mm]{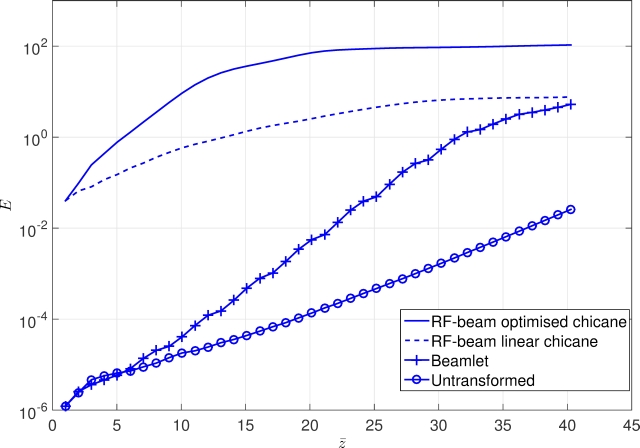}}
\caption{The total radiation field energy $E(\bar{z})$ of the normal (untransformed, no chicanes) FEL, beamlet FEL, and rectangular beams with linear and optimised chicanes.  The radiation is filtered around the resonant frequency, $0.5<\omega / \omega_r < 1.5$.   \label{COMPARISON FIGURE OF ALL SIMS}}
\end{figure}

A comparison of a normal (untransformed beam) FEL amplifier with the methods of beamlets of the previous Section and that of the fourier synthesised rectangular beam of this section  is given in figure~\ref{COMPARISON FIGURE OF ALL SIMS} which plots how the scaled energy $E$ of the radiation pulses evolves with the interaction length $\bar{z}$, where:
\begin{align}
E(\bar{z}) =\int_{-\infty}^{+\infty} |A(\bar{z},\bar{z}_2)| d\bar{z}_2. \label{defE}
\end{align}
Before performing the integral in~(\ref{defE}) the field was first fourier bandpass filtered so that only contributions about resonance in the interval $0.5<\omega / \omega_r < 1.5$ are considered (this removes the significant low-frequency CSE content.) The introduction of the phase=space transform to generate electron beamlets  is seen to increase the exponential growth rate over the normal FEL interaction by a factor of approximately two. While the rectangular electron beams are seen not to have an exponential gain, it is essentially a Coherent Spontaneous Emission process, the starting powers are much greater than the FEL processes which start from spontaneous shot noise. It should be noted that when the CSE simulations predict radiation powers that are a significant fraction of the electron beam energy, that the effects of photon recoil should be included in the model. These effects are not included in the classical simulations presented here.

\section{Conclusion}
This paper has sought to demonstrate what may be possible when electron beams are transformed to alter their properties before injection into an FEL-type system. It is stressed that the methods demonstrated here are not proposals for any specific design or operational wavelength. Rather, they are used to demonstrate possible research directions towards future light sources, some of which have already been envisaged~\cite{beam by design}. 

Here, the focus was to generate  significant radiation output from electron beams that have insufficient beam quality to lase under normal FEL operation. These methods may be developed further and made more specific e.g. to the electron beams generated from plasma accelerator sources which, to date, tend to have relatively high energy spreads. Other possibilities, such as multiple frequency generation, ultra-short pulses, chirped pulses (possibly shorter wavelengths) and others, are potential research areas. One topic that is apparent, but has not been explored here, is the introduction of tapered undulators into the design process.  For example, the introduction of tapered undulators, matched to compensate for the chirped beamlets of above, instead of using chicanes, can be expected to produce interesting radiation output. 

It is noted that the simulations presented here cannot be modelled effectively, or at all, using simulation codes that are used to successfully model the `normal' types of FEL interactions. Unaveraged FEL codes, such as the PUFFIN code used here, are required.

\ack
We gratefully acknowledge support of Science and Technology Facilities Council Agreement Number 4163192 Release \#3; 
ARCHIE-WeSt HPC, EPSRC grant EP/K000586/1;
EPSRC Grant EP/M011607/1; and
John von Neumann Institute for Computing (NIC) on JUROPA at Jlich Supercomputing Centre (JSC), under project HHH20


\appendix
\section*{Appendix}
The final distribution function of a triple-modulator-chicane scheme is given below.  
\begin{footnotesize}
\begin{align}
\small
f(\bar{z}_2,\gamma) =  \frac{1}{2 \pi} \frac{1}{\sigma_\gamma} 
 \frac{1}{\sigma_{\bar{z}_2}}\exp \Bigg(  \Bigg. \frac{-1}{2 \sigma_\gamma^2}     \bigg(  \bigg. [((\gamma+ \Delta \gamma_3 \sin(\frac{1}{2 n_3 \rho}(\bar{z}_2+2D_3(\gamma-\gamma_r) /\gamma_r)+ \phi_3)) \nonumber \\
  +\Delta \gamma_2 \sin(\frac{1}{2 n_2 \rho}((\bar{z}_2+2D_3(\gamma - \gamma_r) /\gamma_r)+ 2D_2((\gamma+\Delta \gamma_3 \sin(\frac{1}{2 n_3 \rho} \nonumber 
 (\bar{z}_2+2D_3(\gamma-\gamma_r) /\gamma_r)+\phi_3))-\gamma_r) /\gamma_r) 
 +\phi_2)) 
 \\ +\Delta \gamma_1 \sin(\frac{1}{2 n_1 \rho}(((\bar{z}_2+2D_3(\gamma-\gamma_r) /\gamma_r)+ \nonumber 
 2D_2((\gamma+\Delta \gamma_3 \sin(\frac{1}{2 n_3 \rho}(\bar{z}_2+2D_3(\gamma-\gamma_r) /\gamma_r)+\phi_3))-\gamma_r) /\gamma_r)  \nonumber \\
 +2D_1(((\gamma+\Delta \gamma_3\sin(\frac{1}{2 n_3 \rho}(\bar{z}_2+2D_3(\gamma-\gamma_r) /\gamma_r)+\phi_3))+ \nonumber 
 \Delta \gamma_2\sin(\frac{1}{2 n_2 \rho}((\bar{z}_2+2D_3(\gamma-\gamma_r) /\gamma_r) \nonumber \\
+ 2D_2((\gamma+\Delta \gamma_3\sin(\frac{1}{2 n_3 \rho}(\bar{z}_2+2D_3(\gamma-\gamma_r) /\gamma_r)+\phi_3))-\gamma_r) /\gamma_r) \nonumber 
 +\phi_2))-\gamma_r) /\gamma_r)+\phi_1)]-\gamma_r \bigg. \bigg) ^2 \Bigg. \Bigg) \nonumber \\
 \exp \Bigg(  \Bigg. \frac{-1}{2 \sigma_{\bar{z}_2}^2} \bigg(  \bigg. [((\bar{z}_2+2  D_3  (\gamma-\gamma_r) /\gamma_r) \nonumber 
 +2  D_2  ((\gamma+\Delta \gamma_3  \sin(\frac{1}{2 n_3 \rho}  (\bar{z}_2+2  D_3  (\gamma- \nonumber
  \gamma_r) /\gamma_r)+\phi_3))-\gamma_r) /\gamma_r) 
  \\ + 2  D_1  (((\gamma+ \nonumber  \Delta \gamma_3  \sin(\frac{1}{2 n_3 \rho}  (\bar{z}_2 + 2  D_3  (\gamma-\gamma_r) /\gamma_r)+\phi_3))+  \nonumber  \Delta \gamma_2  \sin(\frac{1}{2 n_2 \rho}  ((\bar{z}_2 + 2  D_3  (\gamma-\gamma_r) /\gamma_r) \nonumber  \\+ 2  D_2
 ((\gamma+\Delta \gamma_3  \sin(\frac{1}{2 n_3 \rho}  \nonumber  (\bar{z}_2 +2  D_3  (\gamma-\gamma_r) /\gamma_r)+\phi_3))\nonumber  -\gamma_r) /\gamma_r)+\phi_2))-\gamma_r) /\gamma_r]- \bar{z}_c \bigg. \bigg) ^2  \Bigg. \Bigg)
\end{align}

\end{footnotesize}
The energy modulation parameters $\Delta \gamma_{1,2,3}$, modulation frequencies $n_{1,2,3} = k_{1,2,3}/k_w$ and modulation  phases $\phi_{1,2,3}$ are associated with first, second and third modulator sections respectively.  Similarly $D_{1,2,3}$ are the dispersion factors for chicane 1,2 and 3. $\sigma_{\gamma,\bar{z}_2}$ is the standard deviations in $\gamma$ and $\bar{z}_2$.  The resonant energy is defined as $\gamma_r = <\gamma>|_{\bar{z}=0}$ and the electron pulse centre is given by $\bar{z}_c$.
\section*{References}



\begin{thebibliography}{99}

\bibitem{Vinko}Vinko S M \textit{et al} 2012 Nature  \textbf{482}  59

\bibitem{Barends}Barends R M T \textit{et al} 2014 Nature  \textbf{505} 244 

\bibitem{Edwards}Edwards G S \textit{et al} 2003 Rev. Sci. Instrum.  \textbf{74}  3207

\bibitem{lcls} Emma P \textit{et al} 2010 Nat. Photonics \textbf{4 } 641

\bibitem{sacla} Ishikawa T \textit{et al} 2012 Nat. Photonics \textbf{6} 540

\bibitem{McNeil xray}M$^{\mathrm c}$Neil B W J  and Thompson N R 2010 Nat. Photonics \textbf{4} 814

\bibitem{beam by design}Hemsing E, Stupakov G, and Xiang D 2014 Rev. Mod. Phys. \textbf{86} 897

\bibitem{2colour}Campbell L T, McNeil B W J and  Reiche S 2014 New J. Phys. \textbf{16} 103019

\bibitem{Couprie}Couprie M E, Loulergue A, Labat M, Lehe R, and Malka V 2014 J. Phys. B: At. Mol. Opt. Phys. \textbf{47}  234001

\bibitem{Nakajima}Nakajima K 2014 High Power Laser Science and Engineering / Volume 2 / December 2014 / e31

\bibitem{Maier}Maier A R \textit{et al} 2012 Phys. Rev. X \textbf{2} 031019 

\bibitem{Schroeder}Huang Z, Ding Y, and Schroeder C B 2012 Phys. Rev. Lett. \textbf{109} 204801

\bibitem{rf-function beams}Hemsing E and Xiang D 2013 Phys. Rev. ST-Accel. Beams \textbf{16} 010706 

\bibitem{Campbell PUFFIN}Campbell L T and M$^{\mathrm c}$Neil B W J 2012 Physics of Plasmas \textbf{19} 093119

\bibitem{italy1}Ronsivalle C {\em et al.} 2014 New J. Phys. \textbf{16} 033018

\bibitem{mode-locked 1}Thomson NR and M$^{\mathrm c}$Neil BWJ 2008 Phys. Rev. Lett.  {\bf 100} 203901

\bibitem{McNeil SACSE}M$^{\mathrm c}$Neil B W J, Robb G R M, and Jaroszynski D A 2000 Nuclear Instruments and Methods in Physics Research A \textbf{445}  72

\bibitem{bnp} Bonifacio R, Pellegrini C and Narducci L M 1984 Opt. Comm. {\bf 50} 373

\bibitem{Stupakov}Xiang D and Stupakov G 2009 Phys. Rev. ST-Accel. Beams \textbf{12} 030702

\bibitem{bmp} Bonifacio R, M$^{\mathrm c}$Neil B W J and Pierini P 1989 Phys. Rev. A {\bf 40} 4467

\bibitem{noise}M$^{\mathrm c}$Neil B W J, Poole M W and Robb G R M 2003 Phys. Rev. ST-Accel. Beams \textbf{6} 070701

\bibitem{Saldin chirp paper} Saldin E, Schneidmiller E, and Yurkov M 2006 Phys. Rev. ST-Accel. Beams \textbf{9}, 050702

\bibitem{fel14}Henderson J R, Campbell L T and M$^{\mathrm c}$Neil B W J 2014 Proceedings of FEL2014, Basel, Switzerland MOC04 303

\bibitem{bprl}Bonifacio R, De Salvo L, Pierini P, Piovella N and Pellegrini C 1994 Phys. Rev. Lett. {\bf 73} 70

\bibitem{JamesHenderson}Henderson J R and M$^{\mathrm c}$Neil B W J, 2012 EPL \textbf{100} 64001

\bibitem{esase} Zholents A A  2005 Phys. Rev. ST-AB \textbf{8} 040701

\bibitem{James Jones} Jackson F, Angal-Kalinin D, Jones J K, and Williams P H 2013 4th International Particle Accelerator Conference (IPAC13)  Shanghai, China WEPWA063 2262













\end{thebibliography}
\end{document}